# Time, Uncertainty and Non-Locality in Quantum Cosmology

**Maurice Passman**
Adaptive Risk Technology, Ltd.,
London, UK
info@ariskt.co.uk

**Philip V. Fellman**
Suffolk University
Boston, MA
Shirogitsune99@yahoo.com

**Jonathan Vos Post**
Computer Futures
Altadena, CA
Jvospost3@gmail.com

In this paper we build on our previous work and the work of Peter Lynds within a Bohmian framework to consider the intervallic structure of thermodynamic reversibility as well as presenting new considerations for the measurement of uncertainty at cosmic scales. In addition we address the fundamental nature of non-locality as an underlying element of cosmological structure.

## 1 Introduction

Recently there has been some discussion as to whether foundations of physics constitutes a proper subject for a complexity conference. We might begin this discussion by arguing that with respect to foundations, it is very likely that real progress in resolving the remaining issues of general relativity, particularly those issues addressed in quantum cosmology will revolve about broadening, redefining and extending our understanding of the foundations of physics. One particularly pertinent example of this kind of extension is demonstrated in Yaneer Bar Yam's 2006 paper "Is a first order space-time theory possible?"[1] Here, not only are fundamentals extended but the possibility of a different and perhaps ultimately more useful framework for understanding gravitation is introduced. Similarly, as we and others, specifically Julian Barbour, Carlo Rovelli and Peter Lynds, have argued previously a new framework for the understanding of time is necessary for the understanding of dynamical processes and the ways in which we may describe observed behavioral regularities, including emergent phenomena. [2][3][4][5][6]. Finally, there is the question of whether the universe itself may be the result of an "emergent" process of selection [7][8][9]

## 1.1 Time

Common to several of these new theories is the argument that time is an endogenously generated variable, perceived as the consequence of the order of events rather than as an independent, continuous physical process. Following Lynds [2], we argue that the notion of time as a "flowing" series of instants is a fundamental misconception which should not have been carried over from classical physics to either modern quantum mechanics or general relativity (where it confounds any consistent concept time as the endogenous product of system action and state and where such an approach may incorrectly lead one to extend a geodesic which is actually inextensible. [10]. We recapitulate some of our earlier arguments in the paragraphs which follow.

Our notions of time, instants and the flow of time most frequently enter our thinking and discourse as the result of our ordinary experience. As Lynds explains this is primarily the neurobiological function of our perception of intervals of relatively short duration as "present" moments in a continuous or "flowing" stream of time [11]. While this may be subjectively satisfying and almost universally0020experienced it introduces fundamental errors into the discourse of physics. Unfortunately, when scientific research attempts to refute such commonly experienced and widely held notions, like those of temporal instants or instantaneous transformation, the exposition is often met with knee-jerk criticism. The Lynds paper in Foundations of Physics Letters [2] was initially rejected by many readers as a case of not understanding differential calculus. On the contrary, the differential calculus is an excellent and useful abstraction, working very much in the same way that "classes of colors" are described as a "logical fiction" in the Problems of Philosophy.[12] In short, it's not that instantaneous transformations are not useful as an approximation of the behavior of physical systems, but at some more fundamental level it becomes important to understand that in the limiting case they are mere approximations and that, in fact, "time does not flow" nor is there any quantizable or otherwise dimensionless, static instant in time. [2][13]:

> Time enters mechanics as a measure of interval, relative to the clock completing the measurement. Conversely, although it is generally not realized, in all cases a time value indicates an interval of time, rather than a precise static instant in time at which the relative position of a body in relative motion or a specific physical magnitude would theoretically be precisely determined. For example, if two separate events are measured to take place at either 1 hour or 10.00 seconds, these two values indicate the events occurred during the time intervals of 1 and 1.99999…hours and 10.00 and 10.0099999…seconds, respectively. If a time measurement is made smaller and more accurate, the value comes closer to an accurate measure of an interval in time and the corresponding parameter and boundary of a specific physical magnitudes potential measurement during that interval, whether it be relative position, momentum, energy or other. Regardless of how small and accurate the value is made however, it cannot indicate a precise static instant in time at which a value would theoretically be precisely determined, because there is not a precise static instant in time underlying a dynamical physical process. If there were, all physical continuity, including motion and variation in all physical magnitudes would not be possible, as they would be frozen static at that precise instant, remaining that way. Subsequently, at no time is the relative position of a body in relative motion or a physical magnitude precisely determined, whether during a measured time interval, however small, or at a precise static instant in time, as at no time is it not constantly changing and undetermined. Thus, it is exactly due to there not being a precise static instant in time underlying a dynamical physical process, and the relative motion of body in relative motion or a physical magnitude not being precisely determined at any time, that motion and variation in physical magnitudes is possible: there is a necessary trade off of all precisely determined physical values at a time, for their continuity through time.

We have discussed this simple, but very counter-intuitive conclusion elsewhere and simply follow our earlier discussion to summarize the most salient points of that discussion, largely just drawing on Lynds own exposition [6][7][10]. The following section explores Lynds view in some detail [2], and one might also wish to keep in mind Julian Barbour's maxim that "had duration been properly studied in classical physics, its disappearance in the conjectured quantum universe would have appeared natural." [3]

As a natural consequence of this, if there is not a precise static instant in time underlying a dynamical physical process, there is no physical progression or flow of time, as without a continuous and chronological progression through definite indivisible instants of time over an extended interval in time, there can be no progression. This may seem somewhat counter-intuitive, but it is exactly what is required by nature to enable time (relative interval as indicated by a clock), motion and the continuity of a physical process to be possible. Intuition also seems to suggest that if there were not a physical progression of time, the entire universe would be frozen motionless at an instant, again as though stuck on pause on a motion screen. But if the universe were frozen static at such a static instant, this would be a precise static instant of time: time would be a physical quantity. Thus, it is then due to natures very exclusion of a time as a fundamental physical quantity, that time as it is measured in physics (relative interval), and as such, motion and physical continuity are indeed possible.

It might also be argued in a more philosophical sense that a general definition of static would entitle a certain physical magnitude as being unchanging for an extended interval of time. But if this is so, how then could time itself be said to be frozen static at a precise instant if to do so also demands it must be unchanging for an extended interval of time? As a general and sensible definition this is no doubt correct, as we live in a world where indeed there is interval in time, and so for a certain physical magnitude to be static and unchanging it would naturally also have to remain so for an extended duration, however short. There is something of a paradox here however. If there were a precise static instant underlying a dynamical physical process, everything, including clocks and watches would also be frozen static and discontinuous, and as such, interval in time would not be possible either. There could be no interval in time for a certain physical magnitude to remain unchanging. Thus this general definition of static breaks down when the notion of static is applied to time itself. We are so then forced to search for a revised definition of static for this special temporal case. This is done by qualifying the use of stasis in this particular circumstance by noting static and unchanging, with static and unchanging as not being over interval, as there could be no interval and nothing could change in the first instance. At the same time however, it should also be enough just to be able to recognize and acknowledge the fault and paradox in the definition when applied to time.

This position reflects that of Barbour et al, particularly with respect to their work on the dynamics of shape and on York scaling [14] and the Lichnerowicz-York equation [15] and [16]. But there is a further, fundamental discourse error in general relativity caused by Minkowski's confounding of the structure of geodesics with the flow of time. In this case, the Minkowski world-line of a particle is represented as existing even when there is no relative motion or change, arguing that even when nothing happens, time still passes. As Lynds, Barbour, Smolin and others argue, (1) time must be derived from behavior within the light cone and is in this sense endogenous to the light cone and cannot be separated from the interaction of particles, forces and fields within the universe and (2) in this context it may be that time is not a proper first order variable of physical theory, but is rather emergent from these reactions. [2][3][4][5][6][7][8][9][10][16][17]. As we have previously argued [10], this choice of categorical structure has left general relativity in a situation where it is largely limited, in at least this aspect, from further development by poor discourse and an outmoded conceptual schema of space-time.

## 2.1 Uncertainty

Lynds' discourse introduces a new and additional uncertainty constraint in the measurement of time. In addition to quantum uncertainty, because actual measurement takes place on an interval, there are always issues of precision and accuracy with respect to temporal measurement. This is a natural consequence of the position that there is no precisely defined static "instant" of time. In our previous work, we have used the Bell-Lynds Metric to explain what this means in a quantitative sense. Bell's original formulation involved applying a stochastic perturbation to small intervallic measurements of time. This operation allows for a more precise statement of duration while allowing for the kind of measurement error explained by Lynds [2][3]. In "Beeables for quantum theory", Bell argues with respect to dynamics that [19]:

> For the time evolution of the state vector we retain the ordinary Schrodinger equation,
>
> $d/dt |t\rangle = iH |t\rangle$ where H is the ordinary Hamiltonian operator. (4)
>
> For the fermion number configuration we prescribe a stochastic development. In a small time interval $dt$ configuration $m$ jumps to configuration $n$ with transition probability
>
> $dt T_{nm}$, (5)      where  (6)   $T_{nm} = \dfrac{J_{nm}}{D}$ ; and
>
> and            $D_m = \sum |\langle mq|t\rangle|^2$     (8)

provided $T_{nm} > 0$ if $J_{nm} \leq 0$ (9)

From (5) the evolution of a probability distribution $P_n$ over configurations $n$ is given by:

$$d/dt P_n = \sum_m (T_{nm}P_m - T_{mn}P_n) \quad (10)$$

Following Bell we do agree that the mathematical consequence of this intervallic interpretation of the Schrodinger equation (using the stochastic perturbation over a small interval as the error term) is:

$$d/dt|<nq|t>|^2 = \sum_{mp} 2\text{Re} <t|nq><nq|-iH|mp><mp|t> \text{ or}$$

$$d/dt D_n = \sum_m (J_{nm} = \sum_m (T_{nm}D_m - T_{mn}D_n)$$

At this point in our previous paper, we diverged significantly from Bell's interpretation because Bell then uses this exposition to set up a cosmological 3-space and 1-time, Hamiltonian and initial state vector |0>. Our differences are complex, insofar as we extend the logic of the 1-time configuration and timelike geodesics taken over an interval to argue that time is not the integral of an infinite number of precise static instants, but more importantly that time is, in fact a second order, rather than first order variable and that it is an endogenous function of the diffeomorphic evolution of state space over the manifold. Similarly, we offer a somewhat different interpretation of the problem of specialness than do other authors. [7][10][20]

### 3.1 The Schrödinger Wave Equation

There is another sense, independent of the arguments above, in which the Schrödinger wave equation requires additional treatment in the context of non-locality. This is also a position which is complementary to our earlier arguments about the "quantum system" and the definition of said system in the context of non-locality experiments (see Appendix II for a more complete discussion)[10].

Wave functions live on configuration space. Schrödinger called this entanglement. The linearity of the Schrödinger equation prevents the wave function from representing reality. If the equation were non-linear (e.g. reduction models) the wave function living on configuration space still could not by itself represent reality in physical space (Bell, 2004) [19].

### 3.2 The Measurement Problem

Here we follow Bohm (1951 and 1980), and Dürr and Teufel, (2009) [21][22][223]. Given a system that can be described by linear combinations of wave functions $\varphi_1$ and $\varphi_2$. We also have a piece of apparatus that, when brought into interaction with the system, measures whether the system has wave function $\varphi_1$ and $\varphi_2$. Measuring means that, next to the 0 pointer position, the apparatus has two pointer positions, 1 and 2, 'described' by the wave functions $\Psi_0, \Psi_1, \Psi_2$ for which:

$$\varphi_i \Psi_0 \xrightarrow{\textit{Schrodinger evolution}} \varphi_i \Psi_i, \quad i = 1,2$$

The wave function has a support in configuration space which corresponds classically to a set of coordinates of particles (which would form a pointer).

For superposition:

$$\varphi = c_1\varphi_1 + c_2\varphi_2, \quad c_1, c_2 \in \mathbb{C}, \quad |c_1|^2 + |c_2|^2 = 1$$

$$\varphi_i \Psi_0 = (c_1\varphi_1 + c_2\varphi_2) \xrightarrow{Schrodinger\ evolution} c_1\varphi_1\Psi_1 + c_2\varphi_2\Psi_2$$

The outcome on the right side does not concur with experience. It shows a 'macroscopic indeterminacy'. For the Schrödinger cat experiment $\varphi_1$ and $\varphi_2$ are the wave functions of the non-decayed and the decayed atom; $\Psi_0$ and $\Psi_1$ are the wave functions for the live cat and $\Psi_2$ is the wave function for the dead cat. Schrödinger says that this is unacceptable. Why? Isn't the apparatus supposed to be the observer? What counts as an observer? Certainly, in this sense, one cannot draw a coherent explanation from Bohr's arguments as indicated by Bell [10].

### 3.2.1 Decoherence

The evolution of :

$$\varphi_i \Psi_0 = (c_1\varphi_1 + c_2\varphi_2) \xrightarrow{Schrodinger\ evolution} c_1\varphi_1\Psi_1 + c_2\varphi_2\Psi_2 \quad \text{is an instance}$$

of decoherence. The apparatus decoheres the superposition $(c_1\varphi_1 + c_2\varphi_2)$ of the system wave function. Decoherence means that it is in a practical sense impossible to get the two wave packets $\varphi_1\Psi_1$ and $\varphi_1\Psi_2$ superimposed in $c_1\varphi_1\Psi_1 + c_2\varphi_2\Psi_2$ to interfere. Decomposition is this practical impossibility – Bell referred this as 'fapp-impossibility' where fapp = for all practical purposes.

## 3.3 Mechanics

Particle motion is guided by the wave function. The physical theory is formulated with the variables $\mathbf{q}_i \in \mathbb{R}^3, \quad i = 1, 2, 3, ...N$, the positions of the $N$ particles that make up the system, and the wave function $\psi(\mathbf{q}_1, ....\mathbf{q}_N)$ on the configuration space of the system. Quantum randomness – Born's statistical law – is explained on the basis of Bolzmann's principles of statistical mechanics. Born's law is not an axiom but a theorem; Born's statistical law concerning $\rho = |\psi|^2$ is that if the wave function is $\psi$ then the particle configuration is $|\psi|^2$-distributed. Applying this to:

$$\varphi_i \Psi_0 = (c_1\varphi_1 + c_2\varphi_2) \xrightarrow{Schrodinger\ evolution} c_1\varphi_1\Psi_1 + c_2\varphi_2\Psi_2 \quad \text{above implies}$$

that the result $i$ comes with probability $|c_i|^2$.

## 3.4 The Double Slit Experiment

Each particle goes either through the upper or through the lower slit. The wave function goes through both slits and forms after the slits a wave function with an interference pattern:

(A)

Close slit 1 open slit 2
Particle goes through slit 2

It arrives at **x** on the plate with probability $|\psi_2(\mathbf{x})|^2$

Where $\psi_2$ is the wave function which passed though slit 2.

(B)

Close slit 2 open slit 1
Particle goes through slit 1

It arrives at **x** on the plate with probability $|\psi_1(\mathbf{x})|^2$

Where $\psi_1$ is the wave function which passed though slit 2.
Both slits
Both slits are open
The particle goes through slit 1 or slit 2

It arrives at **x** on the plate with probability $|\psi_1(\mathbf{x})|^2 + |\psi_2(\mathbf{x})|^2$

In general:

$$|\psi_1(\mathbf{x}) + \psi_2(\mathbf{x})|^2 = |\psi_1(\mathbf{x})|^2 + |\psi_2(\mathbf{x})|^2 + 2\Re\,\psi_1^*(\mathbf{x})\psi_2(\mathbf{x}) \neq |\psi_1(\mathbf{x})|^2 + |\psi_2(\mathbf{x})|^2$$

Here $\Re$ denotes the real part of a complex quantity. The inequality comes from the interference of the wave packets $\psi_1$, $\psi_2$ which passes through slit 1 and 2. Situations 'Particle goes through slit 2' and 'Particle goes through slit 1' are exclusive alternatives entering 'The particle goes through slit 1 or slit 2', but the probabilities $|\psi_2(\mathbf{x})|^2$ and $|\psi_1(\mathbf{x})|^2$ do not add up – this is because 'Close slit 1 open slit 2', Close slit 1 open slit 1' and 'Both slits are open' are *physically distinct*.

### 3.4.1 Causality, Determinism and Ontology

This type of methodology is often said to aim at restoring determinism to the quantum world. Determinism has nothing to do with ontology. This type of QM is deterministic - but is not an ontological necessity.

## 3.5 Locality

Einstein deduced from Maxwell's equations that space and time change in a different way from Galilean physics when one changes between frames moving with respect to each other. The nature of this change is governed by the unchanging velocity of light when moving from one frame to another. This led to Minkowski showing that a particle needs a position in time and space for its specification therefore implying that a particle in relativistic space should have time and space coordinates.

In this section, we explore the argument that any theory must be nonlocal and attempt to present a mathematical proof to that effect. Nonlocality is crudely defined as meaning that the theory contains action at a distance in the true meaning of the words i.e. faster than light action between separated events.

Action at a distance is such that no information can be sent with superluminal speed – therefore, there is no inconsistency with special relativity. Nonlocality is encoded in the wave function that lives on configuration space and is by its very nature a nonlocal agent. All particles are guided simultaneously by the wave function and *f* the wave function is entangled, the nonlocal action does not get small with particles.

In a two particle system with coordinates $\mathbf{X}_1(t), \mathbf{X}_2(t)$ we have:

$$\dot{\mathbf{X}}_1(t) = \frac{\hbar}{m_1} \Im \left[ \frac{\left. \frac{\partial}{\partial \mathbf{x}} \psi(\mathbf{x}, \mathbf{X}_2(t)) \right|_{\mathbf{x} = \mathbf{X}_1(t)}}{\psi(\mathbf{X}_1(t), \mathbf{X}_2(t))} \right]$$

Therefore the velocity of $\mathbf{X}_1(t)$ at time *t* depends in general on $\mathbf{X}_2(t)$ at time t, no matter how far apart the positions are. In general here means that the wave function is entangled and not a product e.g.: $\psi(\mathbf{x}, \mathbf{y}) = \varphi(\mathbf{x})\Phi(\mathbf{y})$. However, there is no immediate reason why the wave function should become a product when **x** and **y** are far apart (although decoherence is always lurking awaiting an opportunity to destroy coherence i.e. produce an effective product structure).

### 4.1 Bell's Theorem – Part I

Is it possible to describe a quantum mechanics that is local? Einstein, Podoloski and Rosen (EPR) thought so. The EPR argument is interesting as it constitutes part of Bell's proof of the Nonlocality of nature. Bell's response is Bell's theorem: nature is nonlocal.

In this first part of the argument on Bell's Theorem, the EPR argument applied to a simplified EPR Gedanken experiment. This approach is based upon the argument that one can prepare a special pair (L, R) of spin ½ particles that fly apart in opposite directions and which behave in the following well determined fashion.

When both particles pass identically oriented Stern-Gerlach magnets, they deflect in exactly opposite direction. If L has **a**-spin +1/2 then R has **a**-spin -1/2 where **a** is the orientation of the magnets. This is true for all directions **a**. The probability for L up, R down is ½. The two particle wave function is called a singlet state and the total spin of this singlet state is zero. Measuring first the **a**-spin on L, we can predict with certainty the result of the measurement of the **a**-spin on R. This is true even of the measurement events L and R are space-like separated.

Suppose that the experiment is arranged in such a way that a light signal cannot communicate the L result to the R particle before the R particle passes SGM-R. Suppose now that 'locality' holds meaning that the spin measurement on one side has no superluminal influence on the result of the spin measurement on the other side. Then we must conclude that the value we predict for the **a**-spin on R is pre-existing. It cannot have been created by the result obtained on L as we assume locality. If the value pre-exists, then that means that it exists even before the decision was taken in which direction **a** the spin on the left is to be measured. The value therefore pre-exists for any direction **a**. This also holds (by symmetry) for the values obtained on L. by locality, therefore we obtain the pre-existing values of spins on either side in any direction.

We collect the pre-existing values in a family of variables $\mathbf{X}_\mathbf{a}^{(L)}, \mathbf{X}_\mathbf{a}^{(R)} \in {-1,1}$ with **a** indexing arbitrary directions and with $\mathbf{X}_\mathbf{a}^{(L)} = \mathbf{X}_\mathbf{a}^{(R)}$. The locality check is now to ask the question if such pre-existing values actually exist.

## 4.2 Bell's Theorem – Part II

There is no way that the variables can reproduce the quantum mechanical correlations. Choose three directions, given by unit vectors **a**, **b**, and **c** and consider the corresponding 6 variables:

$\mathbf{X}_\mathbf{y}^{(L)}, \mathbf{X}_\mathbf{z}^{(R)}, \mathbf{y}, \mathbf{z} \in {\mathbf{a,b,c}}$. They must satisfy:

${\mathbf{X}_\mathbf{a}^{(L)}, \mathbf{X}_\mathbf{b}^{(L)}, \mathbf{X}_\mathbf{c}^{(L)}} = {-\mathbf{X}_\mathbf{a}^{(R)}, -\mathbf{X}_\mathbf{b}^{(R)}, -\mathbf{X}_\mathbf{c}^{(R)}}$

We wish to reproduce the relative frequencies of the anticorrelation events:

$\mathbf{X}_\mathbf{a}^{(L)} = -\mathbf{X}_\mathbf{b}^{(R)}, \quad \mathbf{X}_\mathbf{b}^{(L)} = -\mathbf{X}_\mathbf{c}^{(R)}, \quad \mathbf{X}_\mathbf{c}^{(L)} = -\mathbf{X}_\mathbf{a}^{(R)}$

Adding the probabilities and using the rules of probability, we get:

$\text{Prob}{\mathbf{X}_\mathbf{a}^{(L)} = -\mathbf{X}_\mathbf{b}^{(R)}} + \text{Prob}{\mathbf{X}_\mathbf{b}^{(L)} = -\mathbf{X}_\mathbf{c}^{(R)}} + \text{Prob}{\mathbf{X}_\mathbf{c}^{(L)} = -\mathbf{X}_\mathbf{a}^{(R)}}$

$= \text{Prob}{\mathbf{X}_\mathbf{a}^{(L)} = \mathbf{X}_\mathbf{b}^{(L)}} + \text{Prob}{\mathbf{X}_\mathbf{b}^{(L)} = \mathbf{X}_\mathbf{c}^{(L)}} + \text{Prob}{\mathbf{X}_\mathbf{c}^{(L)} = \mathbf{X}_\mathbf{a}^{(L)}}$

$\geq \text{Prob}{\mathbf{X}_\mathbf{a}^{(L)} = \mathbf{X}_\mathbf{b}^{(L)} \text{ or } \mathbf{X}_\mathbf{b}^{(L)} = \mathbf{X}_\mathbf{c}^{(L)} \text{ or } \mathbf{X}_\mathbf{c}^{(L)} = \mathbf{X}_\mathbf{a}^{(L)}}$

$= \text{Prob (sure event)} = 1$, as $\mathbf{X}_\mathbf{y}^{(i)}, i = L, R, \mathbf{y} \in {\mathbf{a,b,c}}$ can only take two values.

This is therefore one version of Bell's inequality:

$\text{Prob}{\mathbf{X}_\mathbf{a}^{(L)} = -\mathbf{X}_\mathbf{b}^{(R)}} + \text{Prob}{\mathbf{X}_\mathbf{b}^{(L)} = -\mathbf{X}_\mathbf{c}^{(R)}} + \text{Prob}{\mathbf{X}_\mathbf{c}^{(L)} = -\mathbf{X}_\mathbf{a}^{(R)}} \geq 1$

The logical structure of Bell's nonlocality argument follows.

Let P be the hypothesis of the existence of pre-existing values $X_{\mathbf{a,b,c}}^{L,R}$ for the spin components relevant to this EPRB experiment. Then:

First part: quantum mechanics + locality -> P

Second part: quantum mechanics -> not P

Conclusion: quantum mechanics -> nonlocality

$$\text{Prob}\left(\mathbf{X}_a^{(L)} = -\mathbf{X}_b^{(R)}\right) + \text{Prob}\left(\mathbf{X}_b^{(L)} = -\mathbf{X}_c^{(R)}\right) + \text{Prob}\left(\mathbf{X}_c^{(L)} = -\mathbf{X}_a^{(R)}\right) \geq 1$$

is violated by experimental evidence.[24][25]
We could therefore write:
First part: experimental facts + locality -> P
Second part: experimental facts -> not P
Conclusion: experimental facts -> nonlocality
Nature is therefore nonlocal.

### 5.1 Conclusion

We have discussed several key foundational issues in theoretical physics, with an eye towards the architecture of quantum cosmology in the preceding paper. As we have argued previously, Lynds' treatment of time as an endogenous variable, measured over intervals rather than instants leads inexorably in the direction of the Lynds Conjecture and his unique solution to the problem of specialness [6][20]. In addition, the Bell-Lynds metric strongly suggests that cosmic uncertainty cannot be merely the sum of all local uncertainties.

Additionally, while there is a deterministic element to quantum mechanics, this element is not one of ontological necessity. This is in contradistinction to Lynds' solution to the "problem of specialness", where one can argue about initial state configurations of the early universe from a position of quasi-ontological necessity.

With respect to non-locality, we show, generally, that non-locality is a fundamental property of the configuration space of the universe, and not, simply a special case of strange behavior at very fine scales. We also address Stephen Hawking's argument that while he does not take a current position on "hidden variables", he believes that in the future (as well as in the development of future cosmological theory) new, non-local, hidden variables will be discovered.

In addition to the formal body of the paper, Appendix I provides a detailed mathematical exposition of a number of issues regarding the quantum system and the "quantum problem" of measurement [10][18] addressing Bell's concerns with respect to Bohr's asserted position on the interpretation of quantum mechanical measurements.